\documentclass[12pt]{article}   	
\usepackage{geometry}                		
\usepackage{graphicx}				
\usepackage{amssymb}
\usepackage{amsthm}
\usepackage{url}
\usepackage[authoryear,sort&compress]{natbib}

\title{Fast generalised linear models\\ by database sampling and one-step polishing}
\author{Thomas Lumley\\ Department of Statistics\\ University of Auckland}

\begin{document}
\def\spacingset#1{\renewcommand{\baselinestretch}%
{#1}\small\normalsize} \spacingset{1}
\maketitle

\begin{abstract}
In this note, I show how to fit a generalised linear model to $N$ observations on $p$ variables stored in a relational database, using one sampling query and one aggregation queries, as long as $N^{\frac{1}{2}+\delta}$ observations can be stored in memory.  The resulting estimator is fully efficient and asymptotically equivalent to the maximum likelihood estimator, and so its variance can be estimated from the Fisher information in the usual way.  A proof-of-concept implementation uses R with MonetDB and with SQLite, and could easily be adapted to other popular databases.  I illustrate the approach with examples of taxi-trip data in New York City and factors related to car colour in New Zealand. 
\end{abstract}

\noindent%
\vfill

\newpage
\spacingset{1.45}

\newtheorem{theorem}{Theorem}

\section{Introduction}
Generalised linear models became one of the basic tools of statistical modelling in part because the Newton--Raphson/Fisher scoring algorithm is easy to implement and behaves well.  In this note, I take advantage of the simple form and good behaviour of the algorithm to propose an implementation for large data sets based on in-core fitting to a small subsample followed by a SQL aggregation query.  I present an implementation in R\citep{r-itself} using MonetDB, a column-store database designed for scientific computing\citep{monetdblite}, and a second implementation using SQLite.   The R packages for both these systems allow for zero-configuration database setup and neither requires interprocess communication; they differ in that {\sf MonetDBLite} directly accesses the R heap and so requires less data copying, and also has an efficient primitive for random subsampling. 

The algorithm has two steps
\begin{enumerate}
\item Extract a random subsample of the data into R and compute the maximum likelihood estimator $\tilde\beta$
\item Evaluate the score function at $\tilde\beta$ on the full data set, and perform a one-step Fisher scoring update
\end{enumerate}

In section~\ref{theory} I describe the computation and the theory behind it in more detail, explaining why the one-step estimator is fully efficient.  Section~\ref{example} gives two examples of the use of the method, with comparison to other approaches.  Section~\ref{summary} discusses  possible extensions beyond simple random sampling and generalised linear models.

\section{Theory and methods}
\label{theory}

Suppose we have a sample of size $N$ and want to estimate a  finite-dimensional parameter $\beta_0$ in a generalised linear model, and that a starting estimator $\tilde\beta$ is available.  It is well known that when $\tilde\beta$ is $\sqrt{N}$-consistent, one step of the Newton--Raphson or Fisher scoring algorithms will give an efficient estimator, one that differs by $o_p(N^{-1/2})$ from the maximum likelihood estimator.   It is less well known that the same is true for a wide range of models when $\tilde\beta-\beta_0=o_p(N^{-1/3})$ \citep{cheng-iterations} and that it is true for many generalised linear models when $\tilde\beta-\beta_0=o_p(N^{-1/4})$. The Appendix gives a proof of the latter case, based on the proof for robust linear-regression $M$-estimators by \citet{robust-onestep}; the key step in the proof is that a first-order Taylor expansion of the score has remainder term $O_p(N\|\tilde\beta-\beta_0\|^2)$, not just $o_p(N\|\tilde\beta-\beta_0\|).$

I take advantage of this result by using the maximum likelihood estimator in a random sample of size $n=N^{5/9}$ as a starting value. With this exponent, when $N$ is one billion $n$ is only 100,000, and a subsample of size $n$ can easily be handled in memory.  It is necessary to be able to take a random sample: in my implementation, I use MonetDB's {\tt SAMPLE n} qualifier to the {\tt SELECT} keyword for returning random subsamples, making this step easy. In many other systems it is possible to use a qualifier of the form {\tt WHERE RAND()< k} or {\tt WHERE RANDOM()< k}, with $k$ chosen to give the correct sampling probability. The {\sf SQLite} implementation uses the latter approach);  the resulting subsample size $n$ will be random, but with standard deviation small enough to still ensure $\tilde\beta-\beta=o_p(N^{-1/4})$.

The second key point is that in a generalised linear model the likelihood score is of the form
$$U(\beta)= \sum_{i=1}^n x_iw_i(\beta)(y_i-\mu_i(\beta)),$$
with $w_i=1$ for the most widely used models (those with the canonical link).  For all the commonly-used choices of link and variance function this can be computed with a single SQL query as long as the exponentiation function is available.  Although not part of the SQL standard, exponentiation is widely supported in relational databases, either as a built-in function or as an add-on.

It is also feasible to also compute the expected Fisher information matrix with a single SQL query involving $p\choose 2$ terms, since it is of the form
$$I(\beta) = \sum_{1=1}^N v_i(\beta)x_ix_i^T,$$
for a scalar $v_i()$. I will write $\hat\beta$ for the one-step estimator using $I_n(\tilde\beta)$ computed from the subsample, and $\hat\beta_{\textrm{full}}$ for the one-step estimator using $I_N(\tilde\beta)$ from the whole sample. It turns out that $\hat\beta_{\textrm{full}}$ and $\hat\beta$ are asymptotically equivalent, as demonstrated theoretically in the Appendix and empirically in Section~\ref{example}, so that computing the information on the full sample is not necessary.

\section{Implementation}

Proof-of-concept code in the form of an R package is at \url{github.com/tslumley/dbglm}.  The {\sf tidypredict}\citep{tidypredict} and {\sf dbplyr} packages\citep{dbplyr} are used to simplify the second aggregation query, but the sampling query, which is not supported by {\sf dbplyr}, is written directly in SQL using the {\sf DBI} interface\citep{DBI}.  

The code has certain limitations for speed and simplicity. First, indicator variables will be defined to represent categorical variables, but only for those levels that are present in the subsample; other values will effectively be combined with the reference category. Second, because different databases use different syntax for pseudorandom number generation, the code works only with the {\sf MonetDBLite} database connectors. It should be straightforward to add other databases.  Third, the code supports only a specific list of generalised linear models and does not allow user-defined link and variance functions.  Adding a specific additional link/variance combination that uses only functions supported by the database should be straightforward. Fourth, the code does not support transformations. Variables corresponding to transformations need to be created before {\tt dbglm} is called.

\section{Example and timings}
\label{example}

All analyses were conducted on a Macbook Air laptop with Intel i7-4650U processor at 1.7GHz, 8GiB of memory, and a solid state drive, running OS X 10.13. 2 using R version 3.4.0.   Instructions for obtaining the data and scripts for analysing it are in the {\sf dblgm} package, which can be obtained from \url{https://github.com/tslumley/dbglm}.

\subsection{Are red cars faster?}

The New Zealand Transport Agency maintains a list of all vehicles (currently licensed or not) in the country and their owners. Information on the vehicles, the ``Fleet Vehicle Statistic'' is available for download. The version used here is current at 2017--11--30. It was downloaded from on 2017--12--14.  There are 5 million records; I selected the 3.3 million "PASSENGER CAR/VAN" records. The recorded number of seats varies from 0 to 999 and the recorded number of axles from 0 to 9; I kept only those with 2--6 seats, and 2 axles, leaving 1.7 million records. 

\begin{table}
\label{red}
\caption{Log odds ratios from logistic regression predicting red colour from vehicle characteristics; data from New Zealand Transport Agency. Analysis of all $N=1726134$ records and three realisations of a one-step estimator starting with a subsample of 2917 records.}
\begin{tabular}{llrrrr}
&& Intercept & Power /1000& Seats & Mass/1000\\
\hline
Full data & $\hat\beta$ & -1.04 & 3.12 & -0.149  &-0.302\\
&SE$\times 100$& 2.9 & 3.3 & 0.59 & 0.42\\
Replicate 1 & $\hat\beta$ & -0.99 & 3.10 & -0.160 & -0.296\\
&SE$\times 100$& 2.4 & 3.3 & 0.49 & 0.46\\
Replicate 2 & $\hat\beta$ &-1.06  & 3.05  & -0.143 & -0.301\\
&SE$\times 100$& 2.8 & 3.5 & 0.58 & 0.41\\
Replicate 3 & $\hat\beta$ & -1.01 & 3.08 & -0.153 & -0.302\\
&SE$\times 100$& 2.7 & 3.3 & 0.54 &  0.42\\
\hline
\end{tabular}
\end{table}

I fitted a logistic regression model with outcome variable 1 if the \verb!basic_colour! variable was {\tt RED} and 0 otherwise, using \verb!power_rating!, \verb!number_of_seats!, and \verb!gross_vehicle_mass! as predictors. Table~\ref{red} shows the results from three realisations of the one-step estimator and from a fit to the whole data.   

We see that cars with higher power are more likely to be red, and that this is not due to red cars being bigger; they have lower mass and fewer seats on average.

The one-step estimator took between 1.3 and 1.4 seconds using {\sf MonetDBLite} and 3.1--3.4s using {\sf RSQLite}.  The full maximum likelihood estimator is feasible with a data set of this size; it took 7.6 seconds, not including time for data transfer from the database. The  \verb!bigglm! from the {\sf biglm} package,\citep{biglm} which computes the full maximum likelihood estimator by reading the data set in chunks within each iteration, took 15s using the {\sf MonetDBLite} package and 26s using {\sf RSQLite}.

\subsection{New York Taxis}

In response to Freedom of Information requests, the New York City Taxi Commission has provided data on taxi trips in New York. Here, I analyse the data from 2016 for traditional `yellow cabs' that are authorised to pick up passengers from the street anywhere in the city. The data were downloaded on 2017--12--14 and --15  from \url{https://s3.amazonaws.com/nyc-tlc/trip+data/yellow_tripdata_2016-XX.csv}, with {\tt XX=01} to {\tt 12}.  There are 131 million records.   I filtered the data to only the trips paid by credit card (because these have tip information), and excluded trips of over 50 miles, leaving 86 million records.

I defined the response variable \verb"bad_tip" to indicate a tip of less than 20\%, and defined \verb"night" to be from 8pm to 4am, and {\tt weekend} to be from 8pm Friday until midnight Sunday. 

Table~\ref{taxis} compares the one-step approach to using {\tt bigglm} with {\sf MonetDBLite}.  A low tip is less likely at night, and for airport flights. Long trips are more likely to attract a low tip, but the relationship is not strong. There is little difference between weekend and weekday trips, or with number of passengers. A low tip is much more likely for `negotiated rate' trips: presumably either the rate includes the tip or the customer believes it should.  In this relatively large data set the agreement between the estimators is mostly good, though the coefficent for rate code 3,   ``Newark'', does differ. 

\begin{table}
\caption{Tipping in NY taxis: logistic regression model for odds of $<20$\% tip}
\label{taxis}
\begin{tabular}{lr@{$\quad$(}r@{, }r@{)$\qquad$}r@{$\quad$(}r@{, }r@{)$\quad$}}
\hline
& \multicolumn{3}{c}{One-step} & \multicolumn{3}{c}{Full}\\
& $\hat\beta$ & 95\% & CI & $\hat\beta$& 95\% & CI\\
\hline
(Intercept) & -0.815 & -0.812 & -0.808 & -0.814 & -0.817 & -0.810 \\ 
 weekend& 0.029 & 0.036 & 0.043 & 0.026 & 0.019 & 0.033 \\ 
 night & -0.209 & -0.206 & -0.203 & -0.205 & -0.208 & -0.201 \\ 
 weekend night  & -0.005 & 0.002 & 0.009 & 0.012 & 0.005 & 0.019 \\ 
 passenger count & -0.008 & -0.007 & -0.005 & -0.009 & -0.010 & -0.007 \\ 
 weekend$\times$passenger count & -0.021 & -0.017 & -0.014 & -0.002 & -0.006 & 0.001 \\ 
 night $\times$passenger count  & -0.007 & -0.005 & -0.004 & -0.004 & -0.006 & -0.002 \\ 
weekend night$\times$passenger count & 0.014 & 0.018 & 0.021 & 0.003 & -0.000 & 0.007 \\ 
  trip distance & 0.033 & 0.033 & 0.033 & 0.033 & 0.033 & 0.034 \\ 
\multicolumn{4}{l}{Compared to standard rate}\\
$\quad$To/from JFK  & -0.266 & -0.262 & -0.259 & -0.268 & -0.272 & -0.264 \\ 
$\quad$To/from Newark & -0.138 & -0.130 & -0.121 & -0.199 & -0.209 & -0.189 \\ 
$\quad$Nassau \& Westchester & 0.002 & 0.021 & 0.040 & -0.095 & -0.117 & -0.073 \\ 
$\quad$Negotiated rate& 1.406 & 1.413 & 1.421 & 1.401 & 1.393 & 1.409 \\ 
   \hline
 \emph{Computation time} &\multicolumn{3}{c}{\em cpu: 430s, elapsed: 340s} &   \multicolumn{3}{c}{\em cpu: 890s, elapsed: 920s}\\
   
\end{tabular}

\end{table}






\section{Discussion}
\label{summary}

If the data can be regarded as in random order, so that the first $n$ observations can be used as the subsample, a true one-pass implementation is possible; however, it is important that the subsample truly be representative. 

In a cluster or cloud context It would clearly be straightforward to parallelise both database queries. Parallelising the aggregation query is likely to be helpful; whether parallelising the sampling query is helpful depends on whether the database already has efficient sequential sampling algorithms. 

The ideal size of the subsample will depend on available memory on the local machine, and on the speed of data transfer compared to in-database computation.  When data transfer is fast, as in my examples, a larger $n$ might have been desirable; if data had to come across a network connection from a separate database server it would be desirable to keep $n$ as small as possible.  

The same general approach can be used for other regression models fitted by maximum likelihood, such as those considered by \citet{vglm}, as long as the derivative of the log likelihood with respect to the linear predictors can be expressed in closed form in SQL. For example, it would be feasible to fit a Weibull model, or a zero-inflated Poisson, but fitting the overdispersion parameter of a negative binomial or the degrees of freedom of a Student's $t$ model would not be feasible since the score for these models involves the digamma function.  
Even when the computations are feasible it may sometimes be necessary to use either two iterations or a larger subsample for some models;  \citet{cheng-iterations} shows that a starting estimator with  $o_p(N^{-1/3})$ error is in general required for a single iteration to give an efficient estimator, so $n$ larger than $N^{2/3}$ would be needed. 

If efficient sampling on the outcome variable or predictors is possible, it may be possible to use a smaller subsample. For example, if $Y$ is a rare outcome with   $M\ll N$ observations having $Y=1$, a sample of $M^{\frac{1}{2}+\delta}$ cases and, say, $5M^{\frac{1}{2}+\delta}$ controls will suffice for the subsample.  Whether this or more sophisticated two-phase sampling strategies are useful in practice will depend on details of the database setup such as the cost of constructing new indexes. 

\appendix

\section{Theorem and proof}

Suppose we are fitting a generalised linear model with regression parameters $\beta$, outcome $Y$, and predictors $X$.  Let $\beta_0$ be the true value of $\beta$. Assume the second partial derivatives of the loglikelihood have uniformly bounded second moments on a compact neighbourhood $K$ of $\beta_0$.  Let $U_N(\beta)$ be the score evaluated at $\beta$ on $N$ observations and $I_N(\beta)$ be the expected Fisher information evaluated at $\beta$ on $N$ observations.  Let $\Delta_3$ be the tensor of third partial derivatives of the log likelihood, and assume its elements
$$(\Delta_3)_{ijk}=\frac{\partial^3}{\partial x_i\partial x_jx\partial _k}\log\ell(Y;X,\beta)$$
 have uniformly bounded second moments on $K$.

\begin{theorem}
Let $n=N^{\frac{1}{2}+\delta}$ for some $\delta\in (0,1/2]$, and let $\tilde\beta$ be the maximum likelihood estimator of $\beta$ on a subsample of size $n$. Define $\tilde I(\tilde\beta)=\frac{N}{n}I_n(\tilde\beta)$, the estimated full-sample information based on the subsample
 The one-step estimators
$$\hat\beta_{\mathrm{full}}= \tilde\beta + I_N(\tilde\beta)U_N(\tilde\beta)$$
and
$$\hat\beta= \tilde\beta + \tilde I(\tilde\beta)U_N(\tilde\beta)$$
are first-order efficient
\end{theorem}

\paragraph{Proof:} The score function at the true parameter value is
$$U_N(\beta_0)=\sum_{i=1}^Nx_iw_i(\beta_0)(y_i-\mu_i(\beta_0)$$
By the mean-value form of Taylor's theorem we have
$$U_N(\beta_0)=U_N(\tilde\beta)+I_N(\tilde\beta)(\tilde\beta-\beta_0)+\Delta_3(\beta^*)(\tilde\beta-\beta_0,\tilde\beta-\beta_0)$$
where $\beta^*$ is on the interval between $\tilde\beta$ and $\beta_0$. With probability 1, $\tilde\beta$ and thus $\beta^*$ is in $K$ for all sufficiently large $n$, so the remainder term is $O_p(Nn^{-1/2}n^{-1/2})=o_p(N^{1/2})$. 
Thus
$$I_N^{-1}(\tilde\beta) U_N(\beta_0) = I^{-1}_N(\tilde\beta)U_N(\tilde\beta)+\tilde\beta-\beta_0+o_p(N^{-1/2})$$

Let $\hat\beta_{MLE}$ be the maximum likelihood estimator. It is a standard result that 
$$\hat\beta_{MLE}=\beta_0+I_N^{-1}(\beta_0) U_N(\beta_0)+o_p(N^{-1/2})$$

So
\begin{eqnarray*}
\hat\beta_{MLE}&=& \tilde\beta+I^{-1}_N(\tilde\beta)U_N(\tilde\beta)+o_p(N^{-1/2})\\
&=& \hat\beta_{\textrm{full}}+o_p(N^{-1/2})
\end{eqnarray*}

Now  let ${\cal I}(\tilde\beta)=E_{X,Y}\left[N^{-1}I_N\right]$ be the expected per-observation information.  By the Central Limit Theorem we have  
$$I_N(\tilde\beta)=I_n(\tilde\beta)+(N-n){\cal I}(\tilde\beta)+O_p((N-n)n^{-1/2}),$$
so 
$$I_N(\tilde\beta) \left(\frac{N}{n}I_n(\tilde\beta)\right)^{-1}=\mathrm{Id}_p+ O_p(n^{-1/2})$$
where $\mathrm{Id}_p$ is the $p\times p$ identity matrix.
We have
\begin{eqnarray*}
\hat\beta-\tilde\beta&=&(\hat\beta_{\textrm{full}}-\tilde\beta)I_N(\tilde\beta)^{-1} \left(\frac{N}{n}I_n(\tilde\beta)\right)\\
&=&(\hat\beta_{\textrm{full}}-\tilde\beta)\left(\mathrm{Id}_p+ O_p(n^{-1/2}\right)\\
&=&(\hat\beta_{\textrm{full}}-\tilde\beta)+ O_p(n^{-1})\\
&=&(\hat\beta_{\textrm{full}}-\tilde\beta)+ o_p(N^{-1/2})
\end{eqnarray*}
so $\hat\beta$ is also asymptotically efficient. \qed

\bibliographystyle{apalike}
\bibliography{fast-glm-note.bib}

\end{document}